\documentclass[preprint]{aastex}
\usepackage{emulateapj5}
\def\stacksymbols #1#2#3#4{\def\theguybelow{#2}
        \def\verticalposition{\lower#3pt}
        \def\spacingwithinsymbol{\baselineskip0pt\lineskip#4pt}
        \mathrel{\mathpalette\intermediary#1}}
\def\intermediary #1#2{\verticalposition\vbox{\spacingwithinsymbol
        \everycr={}\tabskip0pt
        \halign{$\mathsurround0pt#1\hfil##\hfil$\crcr#2\crcr
                \theguybelow\crcr}}}
\def\lta{\stacksymbols{<}{\sim}{2.5}{.2}}
\def\gta{\stacksymbols{>}{\sim}{3}{.5}}

\begin{document}

\title{SPATIAL DIFFUSION OF X-RAY EMISSION LINES 
IN THE M87 COOLING FLOW; EVIDENCE FOR 
ABSORPTION}

\author{William G. Mathews$^1$, 
David A. Buote$^{1,2}$\altaffilmark{,4}, 
and Fabrizio Brighenti$^{1,3}$}

\affil{$^1$University of California Observatories/Lick Observatory,
Board of Studies in Astronomy and Astrophysics,
University of California, Santa Cruz, CA 95064\\
mathews@ucolick.org}

\affil{$^2$Department of Physics and Astronomy, 
University of California,
Irvine, CA 92697\\
buote@uci.edu}

\affil{$^3$Dipartimento di Astronomia,
Universit\`a di Bologna,
via Ranzani 1,
Bologna 40127, Italy\\
brighenti@bo.astro.it}

\altaffiltext{4}{Chandra Fellow.}

\begin{abstract}
Recent XMM-Newton observations of 
the cooling flow in M87 indicate 
sharply decreasing oxygen, iron and silicon abundances 
within $\sim 5$ kpc of the galactic center.
This result is unexpected since stellar mass loss and 
Type Ia supernovae are expected to produce 
pronounced central abundance maxima for all 
three elements.
However, it has been suggested that  
many of the strong X-ray lines are optically 
thick and diffuse to larger radii in the cooling flow 
before escaping, falsifying the central abundances.
We verify with radiation transfer calculations 
that this effect 
does indeed occur in the M87 cooling flow, 
but that it is insufficient 
to account for the M87 observations. 
We suggest that some source of continuous 
opacity is required to reduce the central 
X-ray line emission, perhaps by warm gas at 
$T \sim 10^5 - 10^6$ K.
The radial surface brightness 
profiles of X-ray resonance lines are also 
sensitive to turbulence in cooling flows  
which reduces the line center optical depths 
considerably. 
Turbulence may provide sufficient energy to 
continuously heat the warm absorbing gas.
\end{abstract}

\keywords{galaxies: elliptical and lenticular --
galaxies: cooling flows --
galaxies: interstellar medium --
X-rays: galaxies --
X-rays: galaxy clusters}


\section{DIFFUSION OF X-RAY LINE PHOTONS IN M87}

This estimate of the diffusion of X-ray line photons 
in M87 has been motivated
by the exceptionally low abundances of iron, silicon
and oxygen within $\sim 5$ kpc of the center of M87
as recently observed by B\"{o}hringer et al (2000) 
with XMM-Newton. 
We explore the possibility that these abundance variations
are artifacts of the outward diffusion of optically thick
X-ray lines as suggested by 
Gil'franov et al. (1987), Tawara et al. (1997), 
Shigeyama (1998) and B\"{o}hringer et al. (2000).
We find that the expected outward migration of line 
photons does occur, but is insufficient to account for 
the central photon depletion observed.
Instead, an additional source of continuous 
absorption is indicated, as suggested by 
Buote (2000), and we speculate about its physical nature.
Before discussing our results,
we begin with a review of the properties 
of the M87 cooling flow and describe our line transfer 
calculation.

\section{VARIATION OF GAS TEMPERATURE AND DENSITY}

The electron density distribution 
in the M87 cooling flow is taken from 
Nulsen and B\"{o}hringer (1995), corrected to distance 
17 Mpc (1' = 4.945 kpc):
\begin{equation}
n_{e,g} = n_0 [1 + (r/r_o)^p]^{-1}
\end{equation}
with $n_0 = 0.120$ cm$^{-3}$, $p = 1.18$ and
$r_o = 0.5r_e$ where  
$r_e = 8.625$ kpc is the effective radius.
According to B\"{o}hringer et al. (2000) these observations
underestimate the central density which should instead
be represented by a $\beta$ model fit:
\begin{equation}
n_{e,\beta} = n_1 [ 1 + (r/r_{\beta})^2]^{-3 \beta/2} 
\end{equation}
with $n_1 = 0.35$ cm$^{-3}$, $\beta = 0.47$ and 
$r_{\beta} = 1.97$ kpc.
This inner density distribution
is based on ROSAT HRI observations reported in B\"{o}hringer (1999).
Since the central X-ray 
emission in M87 is known to be partly
non-thermal (and time variable), this 
latter density determination may be uncertain.
Nevertheless, 
in the following we assume that $n_e(r)$ is given 
by the maximum of $n_{e,g}(r)$ and 
$n_{e,\beta}(r)$ which intersect at 11.9 kpc.

The (single-phase) gas temperature distribution is found 
by combining the results of B\"{o}hringer et al. (2000) 
within 65 kpc with more distant temperatures measured 
by Nulsen \& B\"{o}hringer (1995).
These results can be fit with the fitting function: 
\begin{equation}
T = T_o \left ( {r_m \over r_{ot}} \right) 
\left[ {r_m \over (r + r_{ot})} 
+ \left( {r \over r_m} \right)^q \right]^{-1}
\end{equation}
with $T_o = 1.5 \times 10^7$ K, $r_m = 5.695$,
$r_{ot} = 1.53$ and $q = 0.22$; all radii are in units of 
$r_e$.
This temperature profile rises from the center toward 
a broad maximum, similar to thermal 
profiles in other well-observed 
bright elliptical galaxy cooling flows 
(Brighenti \& Mathews 1997).
The location of the temperature maximum in M87 
is at an unusually large radius presumably due to the 
relatively high ambient temperature (2 - 3 keV) in 
the Virgo cluster. 
The central cooling of the gas can be understood
as the mixing of hot inflowing gas from the Virgo cluster 
with gas lost from evolving 
stars in M87 which orbit with a lower 
virial temperature than that of the surrounding cluster
(Brighenti \& Mathews 1999a).
The characteristic deep central temperature minimum in M87 
indicates that most of the gas in M87 has not been 
substantially heated by the jet and radio source.

\section{ABUNDANCES, LINE EMISSION AND ABSORPTION}

We assume meteoritic solar abundances with 
$\log(a_E) \equiv \log(n_E/n_H) = -3.07$ and 
$-4.45$ for O and Si respectively.
The radial abundance variation of each element 
in the hot gas relative to solar meteoritic 
is represented with $A_E(r)$.
The line emissivity is assumed to vary with local 
temperature according to the emission coefficients
$\Lambda_{\ell}$ determined by 
Landini \& Monsignori Fossi (L-MF; 1990):
$$
\varepsilon_{\ell} = A_E(r) n_e n_H \alpha \Lambda_{\ell}(T)
~~~(\rm{erg~cm}^{-3}~\rm{s}^{-1})$$
where $\alpha$ is a factor of order unity that 
corrects L-MF abundances to solar meteoritic
and $n_H = n_e (4 - 3\mu)/(2 + \mu)$ is the 
local proton density in a plasma with molecular
weight $\mu = 0.61$.

The emission lines that we consider arise from 
the ground state and are Doppler broadened by thermal 
and turbulent velocities.
The absorption coefficient is 
$$
\kappa (x) = n_H x_i a_E A_E
\left( {\pi e^2 \over m_e c} f_{gu} 
{1 \over \pi^{1/2}} { 1 \over 
\Delta \nu_D} \right) e^{-x^2}
~~~(\rm{cm}^{-1}).$$
Here $a_E A_E(r)$ is the local abundance of the
line-emitting element and $x_i(r)$ is the fractional 
abundance of 
the line-emitting ion interpolated from the results of
Sutherland \& Dopita (1993) for collisional
ionization equilibrium.
Photons are assumed to be emitted in a Gaussian 
profile centered on the line energy $E_o = h \nu_o$
and 
$$ x = { \nu - \nu_o \over \Delta \nu_D}$$
is the dimensionless line frequency with 
\begin{equation}
\Delta \nu_D = {\nu_o \over c}
\left( {2 k T \over A_{Ewt} m_p} 
+ v_{turb}^2 \right)^{1/2}
\end{equation}
where $A_{Ewt}$ is the atomic weight of element E.
The line center optical depth in the radial direction 
is found by integrating $d \tau_o = \kappa(0) dr$.

We are taking $x_i(T)$ and $\Lambda_{\ell}(T)$ 
from two different sources.
However, the error from this possible inconsistency 
is unlikely to be large since collisional ionization
fractions $x_i(T)$ found by most authors, including 
L-MF, are in good agreement 
with those of Arnaud \& Rothenflug (1985).

\section{MONTE CARLO PROCEDURE}

To estimate the diffusion of X-ray line photons, 
we represent the inner cooling flow of M87 with 
$N = 50$ radial zones, each having 
different but individually uniform temperature and
density.
The zone spacing $r_n$ is chosen so that the line luminosity 
is the same from each zone.
The Monte Carlo process begins by choosing 
the zone of origin of each photon based on 
equal weighting among the zones.
The exact birth radius $r_b$ of each photon 
within the chosen zone is then found from 
$$
r_b =[ r_n^3  +  {\cal R} (r_{n+1}^3 - r_n^3)]^{1/3}
$$
where ${\cal R}$ is a random number between 0 and 1.
The initial direction of the photon is
$w = \cos \theta = 2{\cal R} - 1$ 
where ${\cal R}$ is another random number 
and $\theta$ is the angle between the photon
velocity and the direction to the center of the 
cooling flow. 
The initial frequency $x$ is chosen by random 
selection from a Gaussian probability density. 
The distance that the photon moves from its 
origin $r_o$ to the local zone boundary $r_z$ is 
\begin{equation}
\ell = w r_o + \sigma 
[r_z^2 - r_o^2 (1 - w^2)]^{1/2}.
\end{equation}
Here $\sigma = 1$ and $r_z = r_{n+1}$ for outward 
moving photons ($w < w_{cr}$) and
$\sigma = -1$ and $r_z = r_{n}$ for inward 
moving photons ($w > w_{cr}$) 
where $w_{cr} = [1 - (r_n/r_o)^2]^{1/2}$.
In the zone of photon birth $r_o = r_b$. 
If the photon continues to move through several 
zones without scattering, 
$w$ and $r_o$ remain constant and 
Equation (5) can be used to find the 
total path length moved $L$.
The distance $\ell$ traversed by the photon 
across each zone must be 
compared to the free path in the zone:
$$
\ell_s = -\ln({\cal R})/\kappa_n(x_n)$$
where $\kappa_n(x_n)$ is the absorption coefficient 
of the photon in zone $n$.
Since $\Delta \nu_D$ varies from zone to zone it is
necessary to rescale the frequency 
to the local zone $n$ 
($r_n < r < r_{n+1}$):
$x_n(r) = x_n(r_b) [\Delta \nu_D(r_b)/
\Delta \nu_D(r)]^{1/2}$.
If $\ell_s > \ell$, the photon moves into the 
next zone, but if $\ell_s < \ell$ the photon 
is scattered at radius 
$$r_s = ( r_o^2 + L_s^2 - 2 r_o L_s w)^{1/2}$$
where $L_s = \sum \ell_n + \ell_s$.
The photon is then 
scattered into a new random direction 
beginning a new trajectory with $r_o = r_s$, 
assuming complete redistribution in the 
Gaussian line profile. 
The scattering is conservative so all photons 
must ultimately flow out beyond the outermost zone.
When the photon reaches the outer boundary 
of the gas $r_t = r_N$, where the optical depths are
very small, it is assumed to escape freely.

\section{RESULTS AND DISCUSSION}

We describe transfer calculations for two lines: 
SiXIV 2.003 keV 
(L$\alpha$) and OVIII 0.652 keV (L$\alpha$), 
both having f-values $f_{gu} = 0.420$.  
Since these resonance 
lines are among the most optically thick lines of 
Si and O at M87 gas temperatures,
the apparent abundance distribution inferred from each
of these lines -- subsequent to 
spatial diffusion -- must 
correspond to the largest possible central abundance 
depletion. 
For each line-emitting element the apparent radial 
abundance profiles observed with XMM-Newton 
can be fit with 
\begin{equation}
A_E^{(ob)}(r) = A_{ob} \left ( {r_{mob} \over r_{otob}} \right)
\left[ {r_{mob} \over (r + r_{otob})}
+ \left( {r \over r_{mob}} \right)^{qob} \right]^{-1}.
\end{equation}
The apparent (meteoritic) abundance dependence 
of Si and O 
are approximated with:
($A_{ob}$, $r_{mob}$, $r_{otob}$, $qob$) =
(0.1 , 7.4, 0.34, 0.7) and
(0.05, 8.0, 0.5, 0.94) respectively; 
$r_{mob}$ and $r_{otob}$ are in kpc.
The line center optical depths in the
three lines evaluated with 
$A_E^{(ob)}(r)$ (Fig. 1) are smaller than 
those suggested by B\"{o}hringer et al (2000)
in part because we use the higher temperatures 
observed instead of their isothermal approximation. 
If the observed abundance maxima 
are due to outward photon diffusion 
by conservative scattering before escape, 
the total X-ray line luminosity 
within some larger radius (where $\tau \ll 1$) is 
unaffected by the radiative transfer.

According to this line diffusion hypothesis,
the true radial abundance profiles 
are expected to decrease monotonically 
as we represent with 
\begin{equation}
A_E(r) = {A_o \over 1 + (r/r_E)^c} + A_{cf}.
\end{equation}
The first term accounts for the
local enrichment by stellar mass loss and
SNIa explosions, and the second term
represents the abundance of cluster cooling flow gas
flowing into M87,
but the parameterized fit is not intended to 
exactly quantify these two enrichment sources. 
The coefficients in Equation (7) are chosen 
so that the total line luminosities within 50 kpc 
are very close to luminosities based on $A_E^{(ob)}$ 
(see below): 
($A_o$, $r_E$(kpc), $c$, $A_{cf}$) = 
(1.0, 35, 1.4, 0.15) and
(0.4, 12, 2, 0.14) for Si and O respectively.
The $A_{cf}$ are chosen 
for agreement with the XMM-Newton abundances at 
$r \gta 30$ kpc; 
$r_E$ is assumed to be similar to  
$r_e = 8.625$ kpc appropriate for a stellar 
contribution; $c$ is chosen for a reasonable fit 
to the XMM-Newton data in $10 \lta r \lta 30$ kpc;
$A_o$ is adjusted until the total luminosity within 
$r = 50$ kpc (where all lines are optically thin) 
matches that of each line computed with $A_E^{(ob)}(r)$.
With these assumed intrinsic 
abundance profiles $A_E(r)$, the total line luminosities 
within 50 kpc exceed the approximate 
observed luminosities of the Si and O lines 
by about 2 and 13 percent respectively. 
We exceed the oxygen line luminosity 
based on $A_O^{(ob)}(r)$ since this 
abundance is less accurately determined by 
the XMM-Newton data.
The cumulative line
center optical depths are somewhat greater
when evaluated with $A_E(r)$ than with 
$A_E^{(ob)}(r)$ as shown in Figure 1.

Monte Carlo 
radiative transfer for each line 
has been calculated with $10^4$ photons, 
50 spatial zones and abundances $A_E(r)$. 
Results are illustrated in Figure 2. 
The light dotted histograms show the 
abundance of each element that would be 
inferred by observing each line in the 
optically thin limit,
$A_E(birth) = A_E(r) N_{ph,b}/\langle N \rangle$, 
where $N_{ph,b}$ is the number of photons created
in each zone and $\langle N \rangle = 200$ is the
mean number of photons per zone.
The adjacent light-line histograms show 
the variation of the 
apparent abundance indicated by each line  
after radiative diffusion,
$A_E(escape) = A_E(r) N_{ph,e}/\langle N \rangle$, 
where $N_{ph,e}$ is the number of photons that
escape from each zone with no further scattering.
$A_E(escape)$  
can be compared directly with the observed
abundances. 

The light solid-line histograms clearly indicate 
central photon depletions
or flattening within $\sim 5$ kpc.
The apparent Si abundance based on the 2.003 keV line
is clearly self-absorbed and matches the observations
quite well at $r \gta 3$ kpc.
But the amount of photon depletion 
with conservative scattering is clearly 
insufficient to explain the observed O abundance. 
It is not possible to deepen the central 
minima by increasing the O abundance $A_O(r)$ 
(and line opacity) in $r \lta 5$ kpc 
since the global line luminosities 
would exceed those observed.
The OVIII line luminosity may already be too high by 
about 13 percent. 
The central oxygen line luminosity (abundance) only 
levels off and does not decrease toward the origin 
as observed.
Evidently line photons are being lost by 
absorption. 
The central abundances of Si and O -- 
$A_E(0) = 1.15$ and 0.54 respectively -- 
cannot be trusted if the line scattering is 
non-conservative. 
Since the continuous opacity in a warm plasma 
with $T_w \lta 10^6$K decreases rapidly 
with photon energy, $\sigma \propto E^{-3}$, 
the OVIII line should suffer $\sim 30$ times 
more absorption than the SiXIV line, 
in qualitative agreement with the profiles in Figure 2.

Shigeyama (1998) noted that if the outward diffusion
of strong X-ray lines is not considered in the
data reduction, the
central gas density can be underestimated
by $\sim 20 - 50$ percent.
To explore this, we repeated the calculations with
$n_{e,\beta}$ increased by 1.5.
Our conclusions are strengthened by this adjustment in
the gas density.
With higher $n_{e,\beta}$ the luminosity of the
lines within 50 kpc can be matched with a slightly
lower central abundance, resulting in 
central minima in $A_E(escape)(r)$ 
even less pronounced than those in Figure 2.

Studies of optical line emission in the central 
regions of galactic and galaxy group cooling flows 
reveal widespread random motions typically 
about $0.2 - 0.4$ of the sound speed in the 
hot gas (Heckman et al. 1989; Caon et al. 2000).
Since small, optically visible line-emitting regions 
at $T \sim 10^4$ K are likely 
to be strongly coupled to the ambient hot gas, 
we infer that the hot gas must 
share the same turbulent velocities.
Turbulence is expected to strongly influence the 
spatial distribution of X-ray resonance lines,
particularly for more massive line-emitting ions 
(Equation 4).
To demonstrate this, we repeated the transfer calculations
for both lines with 
$v_{turb}^2 = {\cal M}_{turb}^2 (5 A_{E} /6 \mu)$
assuming a turbulent Mach number ${\cal M}_{turb} = 0.3$.
In the presence of turbulence, 
which we regard as more realistic, 
the line center optical depths are greatly 
reduced (Fig. 1), there is 
less outward diffusion (Fig. 2), 
and the apparent central depletion of silicon photons 
is significantly reduced.

We have also computed line profiles for the 
FeXXV 6.69 keV (He 4) resonance line discussed by B\"{o}hringer
et al. but have not discussed its profile here 
because emission in 
this line is unlikely to be representative of 
the intrinsic iron abundance. 
The FeXXV 6.69 keV line is very sensitive to 
the (low) cooling flow temperatures 
$T(r)$ in $r \lta 10$ kpc 
where the line transfer is most relevant 
(e.g. Fig. 2 of Buote, Canizares \& Fabian 1999) and 
relatively few FeXXV line photons are created in 
this low temperature region. 
Because of this, the $\chi^2$ iron abundance determination 
in M87 by B\"{o}hringer et al. 
is dominated by the multitude of 
strong Fe L lines near $\sim 1$ keV.
The many lines that constitute the 
Fe L feature have different atomic 
parameters, so no single L line can 
faithfully represent the iron abundance, 
as we have assumed for the O and Si 
Lyman-$\alpha$ lines.
Since the continuous plasma opacity 
varies as $\sigma \propto E^{-3}$,
the opacity at the FeXXV 6.69 keV is $\sim 1000$ times less 
than the opacity at the OVIII 0.652 keV line. 
An observation of the equivalent 
width of the essentially unabsorbed 
FeXXV 6.69 keV line with projected radius in M87 
would allow a calculation of 
the intrinsic iron abundance distribution, provided 
$T(r)$ is sufficiently well determined.

We conclude that the central minima in apparent abundances and line
luminosities in M87 cannot be understood solely in terms of
conservative photon scattering.  
An additional source of continuous
opacity seems to be required.  
We speculate that the source of this
opacity is the warm absorbing gas at $T \sim 10^5 - 10^6$ K 
as suggested for
M87 by Buote (2000).  
Buote's conclusion is derived from modeling the
ROSAT PSPC data of M87 with coronal plasma emission modified by a
single absorption edge and Galactic absorption. 
The rest energy of the
edge is determined to lie at $\sim 0.6$ keV, near the position of the
OVIII L$\alpha$ line. 
The absorption derived from the ROSAT data 
($\tau \sim 0.5-1$) is spatially 
more extended than that required 
to account for central minima in the X-ray resonance lines.
Because of the limited energy resolution of
ROSAT, the effect of using a simple absorption edge cannot be
distinguished from a model in which the O abundance is allowed to take
low values as done by B\"{o}hringer et al (2000).  

The central
declines in Fe and Si abundances in the B\"{o}hringer et al. data are
difficult to understand since we expect hot gas closer to the
galactic core to be more enriched by centrally peaked stellar 
mass loss and Type Ia supernovae 
(Brighenti \& Mathews 1999b).
Since the oxygen abundance found by B\"{o}hringer et al. 
has a generally 
negative gradient beyond $\sim r_e$, it is apparent 
that the stellar and Type Ia ejecta 
are more oxygen-rich than 
cooling flow gas entering M87 
from larger radii, so a central peak in the O abundance 
would also be expected.
The need for an additional source of continuous
opacity in our line transfer calculations lends new support to the
absorption interpretation.  
If this absorption arises from warm gas,
then its mass is similar to that of the hot gas (Buote 2000).  

The energy source that heats the warm gas is of great interest.
While some of the cooling flow gas may be heated by 
the M87 jet, this region is avoided in the B\"{o}hringer et al.
observations; the deep central minimum in their $T(r)$ shows 
no evidence of AGN heating.
The most likely energy 
source for heating the warm gas is the turbulent motion of 
the hot gas as indicated by the dispersion velocity of 
optical emission lines $v_t \sim 200$ km s$^{-1}$ 
in M87 (Heckman et al. 1989).
These velocities are subsonic in the hot gas, but 
would be supersonic in the warm, X-ray absorbing gas. 
Colliding regions of warm gas would be shock-heated at 
at a rate $H = f n_c m_p v_t^3/r_c$
ergs cm$^{-3}$ s$^{-1}$ where $f$ is the filling factor 
of warm gas clouds of mean radius $r_c$ and electron density
$n_c$. 
Although we expect that some net cooling occurs, 
if shock heating balances radiative cooling 
$C = n_c^2 \Lambda$ ergs cm$^{-3}$ s$^{-1}$
then the cloud size can be estimated: 
$r_c \sim f m_p v_t^3/n_c \Lambda \sim 400 f $ pc
for $v_t = 200$ km s$^{-1}$, $n_c = 0.1$ cm$^{-3}$ 
and cooling coefficient 
$\Lambda = 10^{-22}$ erg cm$^3$ s$^{-1}$. 
With $f \sim 0.5$, this cloud size is consistent with 
the scale of irregularities observed in the optical line
emission and the relative uniformity of the warm plasma 
absorption.

Finally, we speculate further that such extended absorbing
regions are generally present in cooling flows and account for the
apparent deficiency of emission lines characteristic of cooling ions
in the centers of these flows (e.g. Peterson et al. 2000; McNamara et
al. 2000).

We thank referee Paul Nulsen for his helpful suggestions.
Studies of the evolution of hot gas in elliptical galaxies
at UC Santa Cruz are supported by
NASA grant NAG 5-3060 and NSF grant
AST-9802994 for which we are very grateful.
DAB was supported by NASA Chandra Fellowship grant 
PF8-10001 awarded by the Smithsonian Astrophysical 
Observatory under contract NAS8-39073.


\vskip.1in
\figcaption[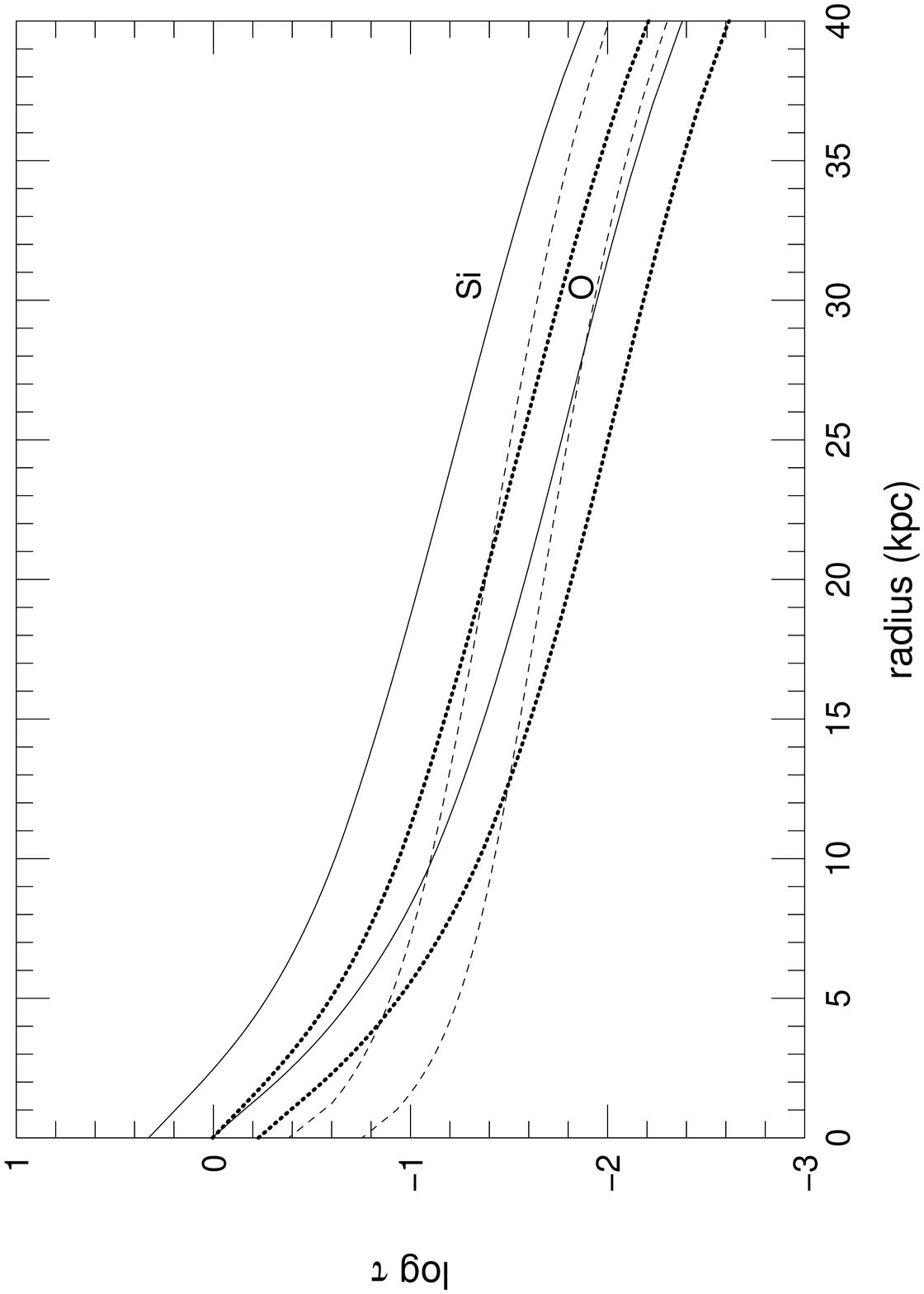]{
Variation of line center optical depths in the 
central cooling flow of M87 in two lines:  
SiXIV 2.003 keV
and OVIII 0.652 keV. The solid and dashed lines 
show optical depths 
resulting from the model $A_E(r)$ 
and directly from the observations $A_E^{(ob)}(r)$ 
respectively.
The heavy dashed lines are line center optical depths 
based on $A_E(r)$ but 
in the presence of turbulence with 
$M_{turb} = 0.3$.
\label{fig1}}

\vskip.1in
\figcaption[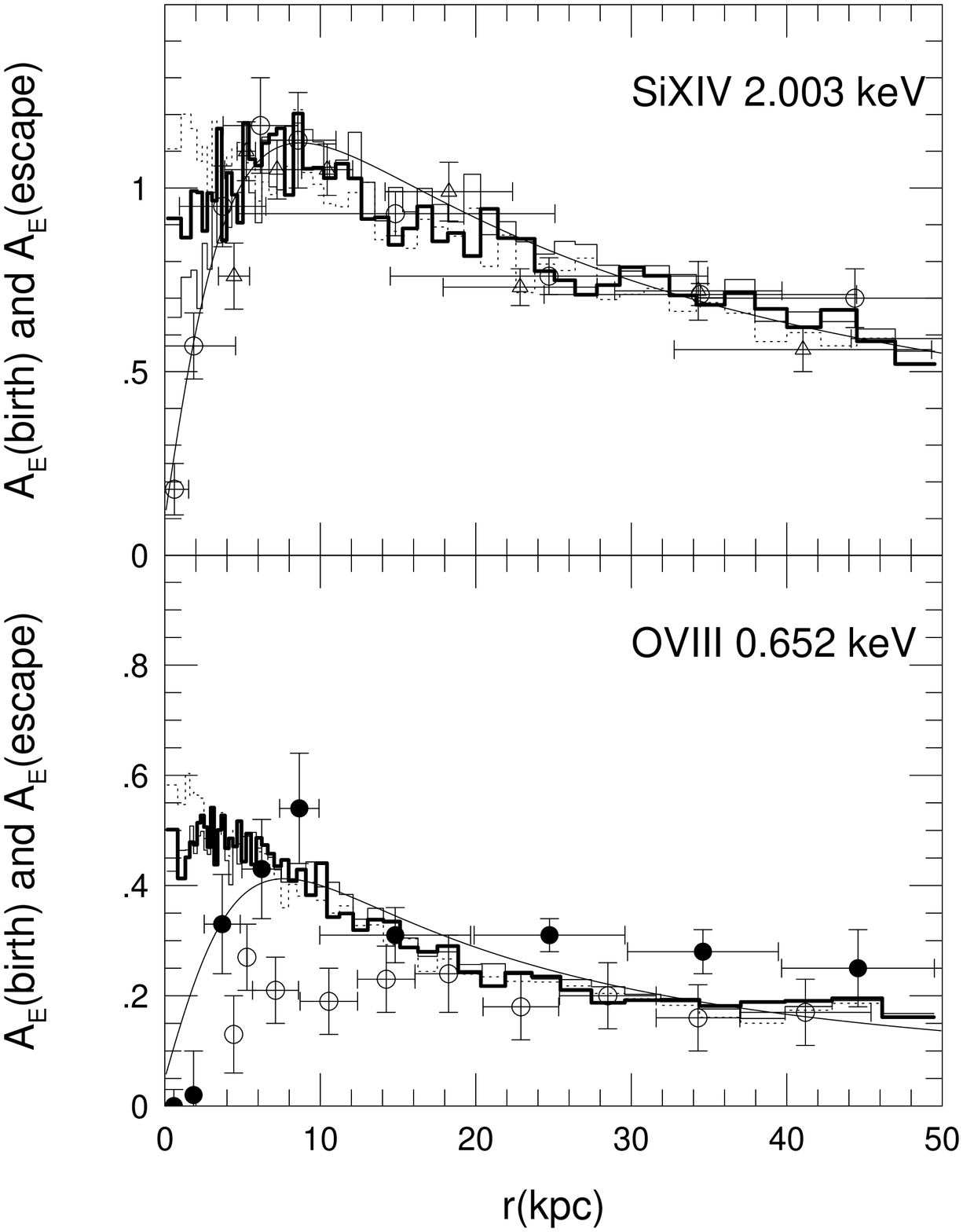]{
Radial variation of silicon and oxygen abundances 
in the M87 cooling flow. 
The observed points including symbol type 
have been taken directly from B\"{o}hringer et al. (2000).
The smooth lines passing through the observations 
are $A_E^{(ob)}(r)$ found from Equation (6).
The dotted histograms show the variation 
of the elemental abundances $A_E(birth)$ 
based on each line using the Monte Carlo program 
in the optically thin limit; apart from statistical 
fluctuations this histogram is identical to $A_E(r)$.
The light solid histograms are the abundance
profiles $A_E(escape)$ 
that would be inferred from each resonance line 
after Monte Carlo spatial diffusion; the 
line center optical depths for these calculations 
are the light solid lines from Figure 1. 
The heavy histograms show the 
effect of turbulence with $M_{turb} = 0.3$.
\label{fig2}}


\begin{references}
\reference{}
Anders, E. \& Grevesse, N. 1989, Geochim. Cosmochim. Acta,
53, 197
\reference{}
Arnaud, M. \& Rothenflug, R. 1985, A\&AS, 60, 425
\reference{}
B\"{o}hringer, H. 1999, in Diffuse Thermal and Relativistic 
Plasma in Galaxy Clusters, H. B\"{o}hringer, L. Feretti, 
P. Schuecker (eds.) MPE Report, p. 115
\reference{}
B\"{o}hringer, H. et al. 2000 A\&A (submitted) (astro-ph/0011459)
\reference{}
Brighenti, F. \& Mathews, W. G. 1997, ApJ, 486, L83
\reference{}
Brighenti, F. \& Mathews, W. G. 1999a, ApJ, 512, 65
\reference{}
Brighenti, F. \& Mathews, W. G. 1999b, ApJ, 515, 542
\reference{}
Buote, D. A. 2000, ApJ (in press) (astro-ph/0008408)
\reference{}
Buote, D. A., Canizares, C. R. \& Fabian, A. C. 1999, 
MNRAS 310, 483
\reference{}
Caon, N., Macchetto, D. \& Pastoriza, M. 2000, ApJS 127, 39
\reference{}
Gil'fanov, M. R., Sunyaev, R. A. \& Churazov, E. M. 
1987, Sov. Astron. Lett., 13, 3 
(Pis'ma Astron. Zh. 13, 7, 1987)
\reference{}
Heckman, T., Baum, S. A., van Breugel, W. J. M. \&
McCarthy, P. 1989, ApJ, 338, 48
\reference{}
Landini, M. \& Monsignori Fossi, B. C. 1990,
A\&AS 82, 229
\reference{}
McNamara, B. R. et al. (2000) in ``Constructing the Universe
with Clusters of Galaxies,'' IAP, Paris (in press) (astro-ph/0012331)
\reference{}
Nulsen, P. E. J. \& B\"{o}hringer, H. 1995, MNRAS 274, 1093
\reference{}
Peterson, J. R. et al. (2000) A\&A (in press) (astro-ph/0010658)
\reference{}
Shigeyama, T. 1998, ApJ, 497, 587
\reference{}
Sutherland, R. S. \& Doptia, M. A. 1993, ApJS, 88, 253
\reference{}
Tawara, Y. et al. 1997, in X-ray Imaging and Spectroscopy
of Cosmic Hot Plasmas, F. Makino \& Mitsuda (eds.),
Universal Academy Press, Tokyo, p. 87
\end{references}
\end{document}